\documentclass[aps,prl,showpacs,twocolumn,floatfix,superscriptaddress,preprintnumbers]{revtex4}
\usepackage{amsmath}
\usepackage{amssymb}
\usepackage{epsfig}
\usepackage{graphicx}
\usepackage{stmaryrd}
\usepackage[dvips]{color}
\def\10{$SO(10)$}
\def\21{SU(2) $\otimes$ U(1) }

\def\422{$SU(4) \otimes SU(2) \otimes SU(2)$}
\def\321{SU(3) $\otimes$ SU(2) $\otimes$ U(1)}
\def\lsim{\raise0.3ex\hbox{$\;<$\kern-0.75em\raise-1.1ex\hbox{$\sim\;$}}}
\def\gsim{\raise0.3ex\hbox{$\;>$\kern-0.75em\raise-1.1ex\hbox{$\sim\;$}}}

\usepackage{dcolumn}% Align table columns on decimal point
\usepackage{bm}% bold math
\usepackage{epsfig}

\newcommand {\ignore}[1]{}

\newcommand{\be}{\begin{equation}}
\newcommand{\ee}{\end{equation}}
\newcommand{\bea}{\begin{eqnarray}}
\newcommand{\eea}{\end{eqnarray}}

\newcommand{\eV}{\mathrm{eV}}
\newcommand{\keV}{\mathrm{keV}}

\newcommand{\GeV}{\mathrm{GeV}}
\def\baselinestretch{1.0}
\def\roughly#1{\mathrel{\raise.3ex\hbox{$#1$\kern-.75em
      \lower1ex\hbox{$\sim$}}}} \def\lsim{\roughly&lt;}
\def\gsim{\roughly&gt;}
\def\e6{E(6)}
\def\321{$SU(3)_{c}\otimes SU(2)_L \otimes U(1)$}
\def\10{SO(10)}

\def\422{SU(4) $\otimes$ SU(2) $\otimes$ SU(2)}

\newcommand{\Gyr}{\mathrm{Gyr}}

\newcommand{\GG}{\Gamma_{18}}

\newcommand{\AHEP}{Instituto de F\'{\i}sica Corpuscular --
  C.S.I.C./Universitat de Val{\`e}ncia \\
  Campus de Paterna, Apt 22085, E--46071 Val{\`e}ncia, Spain}
\newcommand{\oxf}{Oxford Astrophysics, Denis Wilkinson 
Building, Keble Road, OX1 3RH, Oxford, UK }
\begin{document}
\preprint{IFIC/06-39}

\title{Decaying warm dark matter and neutrino masses}
\author{M. Lattanzi}
\email{mxl@astro.ox.ac.uk}
\affiliation{\oxf}
\author{J. W. F. Valle}
\email{valle@ific.uv.es}
\homepage{http://ahep.uv.es}
\affiliation{\AHEP}

\date{\today} % It is always \today, but any date may be explicitly specified

\begin{abstract} 

  Neutrino masses may arise from spontaneous breaking of ungauged
  lepton number.
  Because of quantum gravity effects the associated Goldstone boson - the
  majoron - will pick up a mass.
  We determine the lifetime and mass required by cosmic microwave
  background observations so that the massive majoron provides the
  observed dark matter of the Universe.
  The majoron decaying dark matter (DDM) scenario fits nicely in models where neutrino masses
  arise {\it a la seesaw}, and may lead to other possible cosmological
  implications.
\end{abstract}
\pacs{95.35.+d, 95.36.+x, 98.65.Dx, 14.60.Pq,  14.60.St, 13.15.+g, 12.60.Fr}

\maketitle

%% \newpage
%%% \baselineskip 24pt

A long-standing challenge in particle cosmology is to elucidate the
nature of dark matter and its origin.
A keV weakly interacting particle could provide a sizeable fraction of
the critical density $\rho_{cr} = 1.88 \times 10^{-29} h^{2}$
$\mathrm{g/cm^{3}}$ and possibly play an important role in structure formation,
since the associated Jeans mass lies in the relevant
range~\cite{Bond:1980ha}.
Although we now know from neutrino oscillation experiments that
neutrinos do have mass~\cite{Maltoni:2004ei}, recent cosmological
data~\cite{Lesgourgues:2006nd} as well as searches for distortions in
beta~\cite{Drexlin:2005zt} and double beta decay
spectra~\cite{Klapdor-Kleingrothaus:2004wj} place a stringent limit on
the absolute scale of neutrino mass that precludes neutrinos from
being viable warm dark matter candidates~\cite{Gelmini:1984pe} and
from playing a {\sl direct} role in structure formation.

If neutrino masses arise from the spontaneous violation of ungauged
lepton number there must exist a pseudoscalar gauge singlet
Nambu-Goldstone boson, the
majoron~\cite{chikashige:1981ui,schechter:1982cv}. This may pick up a
mass from non-perturbative gravitational effects that explicitly break
global symmetries~\cite{Coleman:1988tj}.
Despite the fact that the majorons produced at the corresponding
spontaneous L--violation phase will decay, mainly to neutrinos, they
could still provide a sizeable fraction of the dark matter in the
Universe since its couplings are rather tiny.

The decaying warm dark matter particle idea is not new in itself.
However, since early attempts~\cite{Berezinsky:1993fm}, there have
been important observational developments which must be taken into
account in order to assess its viability. 
Specially relevant are the recent cosmological microwave observations
from the Wilkinson Microwave Anisotropy Probe
(WMAP)~\cite{Spergel:2006hy}. For definiteness here we adopt the very
popular possibility that neutrino masses arise {\sl a la
  seesaw}~\cite{Valle:2006vb}.

%{\sl Cosmological DDM scenario}
%

In the following we consider the majoron decaying dark matter (DDM)
idea in a modified $\Lambda$CDM cosmological model in which the dark
matter particle is identified with the weakly interacting majoron $J$
with mass in the keV range. The majoron is not stable but decays non
radiatively with a small decay rate $\Gamma$.  In this DDM scenario,
the anisotropies of the cosmic microwave background (CMB) can be used
to constrain the lifetime $\tau=\Gamma^{-1}$ and the present abundance
$\Omega_{J}$ of the majoron; here we show that the cosmological
constraints on DDM majorons not only can be fulfilled but also can
easily fit into a comprehensive global picture for neutrino mass
generation with spontaneous violation of lepton number.

Although majorons could result from a phase transition, we first
consider them to be produced thermally, in equilibrium with photons in
the early Universe. In this case the majoron abundance $n_J$ at the
present time $t_0$ will be, owing to entropy conservation and taking
into account their finite lifetime:
\begin{equation}
\frac{n_J(t_0)}{n_\gamma(t_0)}=
\frac{43/11}{N_D}\frac{n_J(t_D)}{n_\gamma(t_D)}\;e^{-t_0/\tau},
\end{equation}
where $t_D$ is the time of majoron decoupling, and $N_D$ denotes the
number of quantum degrees of freedom at that time.  If $T(t_D)\gtrsim
170\,\GeV$, then $N_D=427/4$ for the particle content of the standard
model. On the other hand, in the context of a supersymmetric extension
of the SM, there would possibly be, at sufficiently early times, about
twice that number of degrees of freedom.  Moreover, just after
decoupling, the majoron to photon ratio $r \equiv
n_J(t_D)/n_\gamma(t_D)$ is equal to 1/2.
The present density parameter of majorons is then
\begin{equation}
\Omega_J h^2=\frac{m_J}{1.25\,\keV}e^{-t_0/\tau},
\end{equation}
where we have used the standard value of $N_D$.

Alternatively, if majorons were produced already out of equilibrium
there is a range of possible models, which we write generically as
\begin{equation}
\Omega_J h^2=\beta \frac{m_J}{1.25\,\keV}e^{-t_0/\tau}.
\label{eq:omj}
\end{equation}
where the quantity $\beta$ parametrizes our ignorance about both the
exact production mechanism, and the exact value of $N_D$. When $\beta
=1$, we recover the scenario described above, with $r=1/2$ and
$N_D=427/4$.

\begin{center}
\begin{figure}[h]
\includegraphics[clip,width=1.\linewidth]{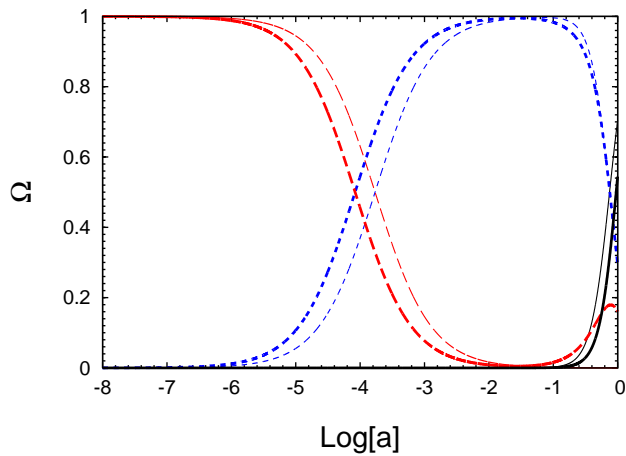}
\caption{(Color online) Evolution of abundances in the standard (thin lines) and DDM
  (thick lines) universe scenario: blue/short dashed, red/long dashed and black/solid correspond to the
  matter, radiation and $\Lambda$ components, respectively.}
\label{fig:lss}
%\vglue -1cm
\end{figure}
\end{center}

Clearly if the majoron is to survive as a dark matter particle it must
be long-lived, $\tau_J \geq t_0$. However, a more stringent bound
follows by studying the effect of a finite majoron lifetime on the
cosmological evolution and in particular on the CMB anisotropy
spectrum. In the DDM scenario, due to particle decays, the dark matter
density is decreasing faster than in the standard cosmological
picture.  This changes the time $t_{eq}$ of radiation-matter equality.
This means that, for a fixed $\Omega_J$, there will be more dark
matter at early times, and the equality will take place earlier, as
illustrated in Fig.~\ref{fig:lss}.
The present amount of dark matter is $\Omega_{DM} = 0.25$ for both
models; $\Gamma^{-1} = 14\,\Gyr$ in the DDM model. Other relevant
parameters are $\omega_b = 2.23\times 10^{-2}$ and $h=0.7$. The time
at which the blue and red lines cross is the time of matter-radiation
equality; for fixed $\Omega_{DM}$, it shifts to earlier times as the
majoron lifetime decreases.

%%%%

The time of matter-radiation equality has a direct effect on the CMB
power spectrum. The gravitational potentials are decaying during the
radiation dominated era; this means that photons will receive an
energy boost after crossing potential wells. This so-called early
integrated Sachs-Wolfe (EISW) effect ceases when matter comes to
dominate the Universe, since the potential are constant during matter
domination. The overall effect is to increase the power around the
first peak as the equality moves to later times.

On the other hand, since $\tau \gtrsim t_0$, we expect the majoron
decays to make the gravitational potentials vary again in the late
stage of the cosmological evolution. This will induce a similar effect
to the one described above, only affecting larger scales due to the
increased horizon size. This late integrated Sachs-Wolfe (LISW) effect
results then in an excess of power at small multipoles.

%%%

Both effects can be used in principle to constrain the majoron
lifetime and cosmological abundance. In order to be quantitative, we
have developed a modified version of the \verb+CAMB+
code~\cite{Lewis:1999bs}, which enables us to compute the CMB
anisotropy spectrum once the majoron lifetime and abundance are given
in addition to the standard $\Lambda$CDM model parameters.

Even if a keV majoron would constitute a warm dark matter particle, it
behaves as cold dark matter insofar as the calculation of its effect
on the CMB spectrum is concerned, since CMB measurements cannot
discriminate between cold and warm dark matter~\cite{Viel:2005qj}. The
latter behaves differently from a cold one on scales smaller than its
free-streaming length $\lambda_{fs}$. For a particle mass in the keV
range, we have $\lambda_{fs}\sim 1\,\mathrm{Mpc}$ which corresponds in
the CMB to a multipole $\ell\sim \textrm{few thousands}$.
The formalism needed to account for the cosmological evolution of an
unstable relic and of its light decay products, has been developed for
example in Ref.  \cite{Kaplinghat:1999xy,Ichiki:2004vi}, including the
modifications in both background quantities and perturbation
evolution.  

Two distinct mechanisms effective at very different times characterize
the effect of DDM on the CMB.
It is therefore convenient to choose a parametrization that can take
advantage of this fact.
% As mentioned above, the effect of DDM on the CMB occurs mainly due to
% two distinct mechanisms effective at very different times. It is
% therefore convenient to choose a parametrization that can take
% advantage of this fact.
%
In particular, the ``natural'' parametrization $(\Omega_J,\,\Gamma)$
has the drawback that both parameters affect the time of
matter-radiation equality.  It is more convenient to define the
quantity
\begin{equation}
Y \equiv \left.\frac{\rho_J}{\rho_b}\right|_{t=t_{\mathrm{early}}},
\end{equation}
where $\rho_b$ is the energy density of baryons, and
$t_{{\mathrm{early}}}\ll t_0 \lesssim \tau$.  As long as this
condition is fulfilled, the value of $Y$ does not depend on the
particular choice of $t_\mathrm{early}$, since the ratio
$\rho_J/\rho_b$ is asymptotically constant at small times. Given that
$t_{eq}\ll\tau$ we can use the value of $Y$ to parametrize the
relative abundance of majorons at matter-radiation equality.
In order to simplify notation let us also define $\GG \equiv
\Gamma/(10^{-18} \mathrm{sec}^{-1})$; in this way, $\GG=1$ corresponds
to a lifetime $\tau \simeq 30\, \Gyr$.

\begin{center}
\begin{figure}[h]
\includegraphics[clip,width=1.\linewidth]{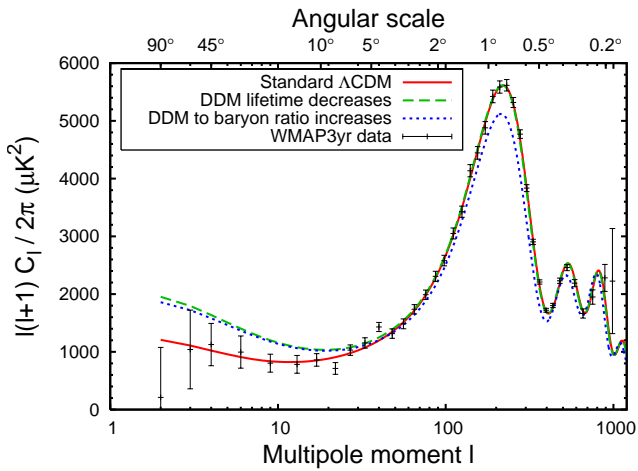}
\caption{(Color online) Effect of DDM parameters on the CMB anisotropy spectrum. The value of the parameters are as follows. Red/solid: fiducial model $(\GG,\,Y) = (0,\,4.7)$. 
Green/dashed: $(\GG,\,Y) = (1.2,\,4.7)$.
Blue/dotted: $(\GG,\,Y) = (1.2,\,5.6)$. See text.
}
\label{fig:cmb}
%\vglue -1cm
\end{figure}
\end{center}

In the parametrization $(Y,\, \Gamma)$, fixing the other parameters,
$Y$ determines the time of equality, while $\Gamma$ mainly affects the
magnitude of the LISW effect.
%Using the parametrization $(Y,\, \Gamma)$ we have that,
%when the other parameters are fixed, $Y$ determines the time of 
%equality, while $\Gamma$ mainly affects the magnitude of the
%LISW effect. 
%The advantage of using the parametrization $(Y,\, \Gamma)$ is that,
%when all other parameters are fixed, the time of matter-radiation
%equality is uniquely determined by $Y$, while the magnitude of the
%LISW effect is largely determined by $\Gamma$.
We show in Fig.
\ref{fig:cmb} how the two physical effects are nicely separated in
this parametrization.
We start from a fiducial model with $\GG = 0$ and $Y = 4.7$; all other
parameters are fixed to their WMAP best-fit values.  The values of
$\GG$ and $Y$ are chosen in such a way to give $\Omega_J h^2 = 0.10$,
so that this fiducial model reproduces exactly the WMAP best-fit.
At a larger majoron decay rate $\GG = 1.2$, i.e., $\Gamma^{-1}\simeq
27\,\Gyr$ the LISW effect makes, as expected, the power at small
multipoles increase, while the shape of the spectrum around the first
peak does not change, since the abundance of matter at early times has
not changed.
Finally, increasing $Y$ by 20\% the height of the first peak decreases
accordingly, while the largest angular scales (small $\ell$) are
nearly unaffected. A small decrease in power in this region is
actually observed, and can be explained by noticing that increasing
the matter content we delay the onset of the $\Lambda$ dominated era,
reducing the $\Lambda$ contribution to the LISW effect.

Another advantage of using the above parametrization is that
$Y$ is directly related to the majoron mass through:
\begin{equation}
Y = 0.71 \times \left(\frac{m_J}{\keV}\right)\,\left(\frac{\beta}{\Omega_b h^2}\right) .
\end{equation}
We are now ready to compute the constraints that CMB observations put
on the majoron abundance and lifetime. 
As seen from Fig.  \ref{fig:cmb}, even a lifetime twice as long as the
present age of the Universe, is quite at variance with respect to the
WMAP data.  
However one must take into account the fact the values of the other
cosmological parameters can be arranged in such a way as to reduce or
even cancel the conflict with observation, i.~e. degeneracies may be
present in parameter space.  
In order to obtain reliable constraints for the majoron mass and
lifetime, we perform a statistical analysis allowing for the variation
of all parameters.  
This is better accomplished using a Markov chain Monte Carlo approach;
we used to this purpose the widely known \verb+COSMOMC+ code
\cite{Lewis:2002ah}.

In our modified flat ($\Omega = 1$) $\Lambda$CDM model, all the dark
matter is composed of majorons. This means that no stable cold dark
matter is present\cite{Hirsch:2004he}.
The 7-dimensional parameter space we explore therefore includes the
two parameters $(Y,\,\Gamma)$ defined above, in addition to the five
standard parameters, namely: the baryon density $\Omega_b h^2$, the
dimensionless Hubble constant $h$, the reionization optical depth
$\tau$, the amplitude $A_s$ and spectral index $n_s$ of the primordial
density fluctuations. 
The cosmological constant density $\Omega_\Lambda$ depends on the
values of the other parameters due to the flatness condition. We
compare our results with the CMB anisotropies observed by the WMAP
experiment.
Once the full probability distribution function for the seven base
parameters has been obtained in this way, the probability densities
for derived parameters, such as the majoron mass $m_J$, is obtained.

\begin{center}
\begin{figure}[h]
\includegraphics[clip,width=0.85\linewidth]{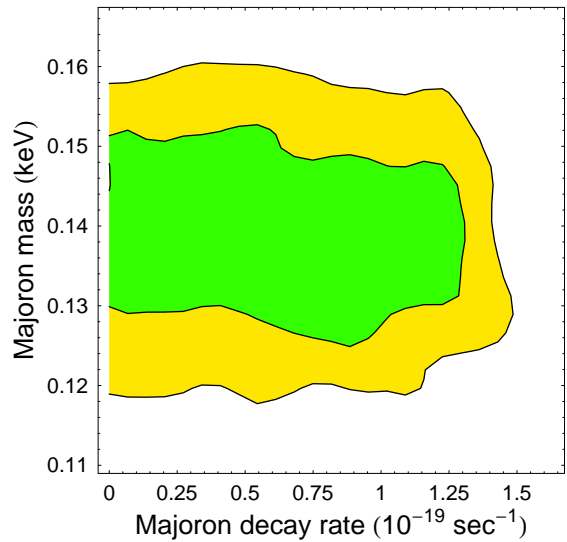}
\caption{(Color online) Contours of the 68\% (green/dark) and 95\% (yellow/light)
  confidence regions in the $(\Gamma_J,\,m_J)$ plane.}
\label{fig:cmb-mjvstau}
%\vglue -0.8cm
\end{figure}
\end{center}

We show our result in Fig.  ~\ref{fig:cmb-mjvstau}, where we give the
68\% and 95\% confidence contours in the $(m_J,\,\Gamma)$ plane, for
the case $\beta=1$, i.e., thermal majoron production and $N_D=427/4$.
We note that these parameters are not degenerate one with the other,
so the respective constraints are independent. Similarly, we find no
degeneracy between $\Gamma_J$ and the five standard parameters. The
marginalized 1-dimensional limits for $\Gamma_J$ and $m_J$ are:
\begin{eqnarray}
\Gamma_J < 1.3\times 10^{-19} \mathrm{sec}^{-1}\\
0.12\, \keV < m_J < 0.17\, \keV
\end{eqnarray}
Expressed in terms of the majoron lifetime our result implies $\tau >
250\, \mathrm{Gyr}$, nearly a factor 20 improvement with respect to
the naive limit $\tau~>~t_0~\simeq~14\,~\mathrm{Gyr}$, illustrating
the power of CMB observations in constraining particle physics
scenarios.
%%%
%
Let us comment on the possibility that $\beta\neq 1$. From eq.
\ref{eq:omj}, it can be seen that this amounts to the transformation
$m_J\to \beta m_J$. For example, as we have already pointed out, $N_D$ 
can be as large as $427/2$, so that
the above limit would read $0.24\,\keV<m_J<0.34\,\keV$. In general, if
we allow for the possibility of extra degrees of freedom in the early
Universe, we always have $\beta<1$ and then
%\vglue -0.5cm
$m_J>0.12\,\keV~. $	
If instead majorons are produced non-thermally, one will in general
have $\beta>1$.

Let us now briefly comment on the particle physics model. The simplest
possibility is that neutrino masses arise {\it a la
  seesaw}~\cite{Valle:2006vb}.  In the basis $\nu, \nu^c$ (where $\nu$
denote ordinary neutrinos, while $\nu^c$ are the \21 singlet
``right-handed'' neutrinos) the full neutrino mass matrix is given as
\begin{equation}
\label{ss-matrix-123} {\mathcal M_\nu} = \left(\begin{array}{cc}
    Y_3 v_3 & Y_\nu v_2 \\
    {Y_\nu}^{T} v_2  & Y_1 v_1 \\
\end{array}\right) 
\end{equation}
and involves, in addition to the singlet, also a Higgs triplet
contribution~\cite{schechter:1980gr} whose vacuum expectation value
obeys a ``vev seesaw'' relation of the type $v_3 v_1 \sim v_2^2$.
The Higgs potential combines spontaneous breaking of lepton number and
of the electroweak symmetry. The properties of the seesaw majoron and
its couplings follow from the symmetry properties of the potential and
were extensively discussed in \cite{schechter:1982cv}.
Here we assume, in addition, that quantum gravity
effects~\cite{Coleman:1988tj} produce non-renormalizable Planck-mass
suppressed terms which explicitly break the global lepton number
symmetry and provide the majoron mass, which we can not reliably
compute, but we assume that it lies in the cosmologically
interesting keV range.

In all of such models the majoron interacts mainly with neutrinos,
proportionally to their mass~\cite{schechter:1982cv}, leading to
\begin{equation}
  \label{eq:jj}
\tau  (J \to \nu \nu) \approx \frac{16\pi}{ m_J} \frac{v_{1}^2}{m_\nu^2}.
\end{equation}
The limits obtained above from the WMAP data can be used to roughly
constrain the lepton number breaking scale as $v_1^2 \gtrsim
3\times\left( 10^6\,\GeV \right)^2$, for $m_\nu \simeq 1\eV$.

The massive majoron has also a sub-leading radiative decay mode, $J \to
\gamma \gamma$, making our DDM scenario potentially testable through
studies of the diffuse photon spectrum in the far ultra violet.  A
more extended investigation of these schemes will be presented
elsewhere including other cosmological data such as the large scale
structure data from the Sloan Digital Sky Survey (SDSS)~\cite{prepa}.
In contrast, we do not expect the data from upcoming CMB experiments
like Planck to substantially improve our bounds on the majoron decay
rate, since they mainly affect the large angular scales where the
error bars have already reached the limit given by cosmic variance.
We also note that direct detection of a keV majoron is possible in a
 suitable underground experiment \cite{Bernabei:2005ca}.
 
{\sl Acknowledgements} Work supported by MEC grant FPA2005-01269, by
EC Contracts RTN network MRTN-CT-2004-503369.  ML is currently
supported by INFN and the University of Rome ``La Sapienza''. The
authors would like to acknowledge S.  Pastor for  computing help.

 \def\baselinestretch{1}%

%  \bibliographystyle{h-physrev4} 
%  \bibliography{cosm-ref-lattanzi,valle-ref}

\end{document}